# On the Interesting World of Fractals and Their Applications to Music

Pabitra Pal Choudhury*, Sk. Sarif Hassan,
Applied Statistics Unit, Indian Statistical Institute, Kolkata, 700108, INDIA
Email: pabitrapalchoudhury@gmail.com, sarimif@gmail.com

Sudhakar Sahoo,
Silicon Institute of Technology, Silicon Hills, Patia, Bhubaneswar, 751024
Email: sudhakar.sahoo@gmail.com

Soubhik Chakraborty
Department of Applied Mathematics
Birla Institute of Technology,
Mesra, Ranchi-835215, INDIA
Email: soubhikc@yahoo.co.in

*Corresponding author

***Abstract*****—** In this paper we have defined one function that has been used to construct different fractals having fractal dimensions between 1.58 and 2. Also, we tried to calculate the amount of increment of fractal dimension in accordance with the base of the number systems. Further, interestingly enough, these very fractals could be a frame of lyrics for the musicians, as we know that the fractal dimension of music is around 1.65 and varies between a high of 1.68 and a low of 1.60. Further, at the end we conjecture that the switching from one music fractal to another is nothing but enhancing a constant amount fractal dimension which might be equivalent to a kind of different sets of musical notes in various orientations.

## I. INTRODUCTION

In this paper an attempt is being made to make an association between natural number and fractals, with the help of Carry Value Transformation (CVT) earlier defined by the same authors [6]. Numbers corresponding to different number systems with base b (b=2, 3, 4…etc) signifies different fractals. With the help of CVT in different base of number system we have generated fractals whose dimension lying in the semi open interval [1.58, 2) [see figure 2 to figure 5]. It should be noted that we are traversing this interval discretely, but very densely also. That is for any given number from [1.58, 2) we could be able to give a fractal whose dimension is nearer to that given number. In our journey we have got a fractal whose dimension is 1.68, fortunately this very fractal dimension is the fractal dimension of music (Sri Lankan, Chariots of Fire) [7]. So this fractal could be a frame of lyrics for music. And undoubtedly there are a lot of fractals, which are of fractal dimension 1.68, and possibly these fractal-frames make new lyrics for music. In general, the fractal dimension of music is around the number 1.65, and our generated fractals could interpolate the number 1.65, as fractal dimension. In this paper, also we have tried to calculate the amount of increment of fractal dimension in accordance with base of the number systems. And in switching of fractals from one base to another, the increment of fractal dimension is constant, which is 1.58.

The organization of the current paper is as follows. In section II some of the basic concepts on fractals, fractal dimension and its relation in music as well as the concept of CVT in binary number system are discussed. In section III, we have generalized the concept of formation of fractals using CVT in any arbitrary bases of the number systems. Finally a conclusion is drawn in section IV.

## II(A) REVIEW OF SOME FUNDAMENTALS OF FRACTALS

The scientific community was very much worried due to their inability to describe the shape of cloud, a mountain, a coastline or a tree on using the traditional Mathematical tools. In Nature, clouds are not really spherical, mountains are not conical, coastlines are not circular, even the lightning doesn't travel in a straight line. More generally, we would be able to conclude that many patterns of nature are so irregular and fragmented, that, compared with *Euclid Geometry* –a term, can be used in this regard to denote all of the standard geometry. Mathematicians have over the years disdained this challenge and have increasingly chosen to flee nature by devising theories unrelated to natural objects we can see or feel. After a long time, responding to this challenge, Benoit Mandelbrot developed a new geometry of nature and implemented its use in a number of diverse arenas of science such as Astronomy, Biology, Mathematics, Physics, and Geography and so on [1-4]. This new-born geometry can describe many of the irregular and fragmented (chaotic) patterns around us, and leads to full-fledged theories, by identifying a family of shapes, now-a-days which we people call 'FRACTALS'.

### *Fractal dimension*

Now let us try to define what fractal dimension (Similarity dimension) is. Given a self-similar structure [figure 2], there is a relation between the reduction factor (scaling factor) 'S' and the number of pieces 'N' into which the structure can be divided; and that relation is as follows…

$N = 1/S^D$, equivalently,

$$\text{i.e. } D = \log(N)/\log(1/S)$$

This 'D' is called the Fractal dimension (Self-similarity dimension).

## II(B) FRACTALS AND MUSIC

A frequently applied form of musical structure generation from the field of chaos theory uses various shapes of what is called fractional noise (also called fractal noise). The term refers to various forms of noise that can be differentiated with respect to their *spectral density* that expresses the distribution of noise power with frequency. The most interesting form which aids in musical structure genesis is *pinc noise*, also referred to as $1/f$ noise, whose behavior lies somewhere between two extremes, namely, *white noise*, with spectral density $1/f^0$ (implying a stochastic process of uncorrelated random variables) and *Brownian noise*, with spectral density $1/f^2$ (implying a stochastic process involving highly correlated random variables). In a musical mapping (for example on pitches) the features of pink noise yield a progression in which stepwise movement and melodic jumps bear a well-balanced relation.

Richard F. Voss and John Clarke [8] described the features of spectral density in recordings of different musical genres and showed their parallels to the peculiarities of $1/f$ noise observing it to be a good choice of stochastic composition. They also extended their one-dimensional model to a two-voice structure which is partly correlated but whose rhythmic shape can be designed using $1/f$ noise. Based on these works, Charles Dodge and Thomas A Jerse [9] described the generation of $1/f$ sequences and produced examples of musical mappings of these different noise forms.

Jeff Pressing [10] mapped the *orbit* (also called *trajectory*: the sequence of values that forms the result for a particular variable of an iterative equation system; these values mostly approach an attractor of a certain shape) of non-linear equation systems (sometimes called non-linear maps) on musical parameters. The map output is used to control pitch, duration, envelope attack time, dynamics, textural density and the time between notes of single events of synthesized sounds. Rick Bidlack [11] also mapped the orbit of two, three and four dimensional equation systems on musical parameters.

Another interesting approach made by Jeremy Leach and John Fitch [12] derives a tree structure from the orbit of a chaotic system, the design being inspired by the works of Lerdahl and Jackendoff (see chap. 4 of [13]). It consists of a hierarchical arrangement of scales and note values and results from the interpretation of the values of the orbit as hierarchic positions of nodes. Concrete note values are produced by interpreting nodes of higher hierarchic order as pitches that structure a melodic progression.

In an early work, Przemyslaw Prusinkiewicz [14] described the simple mapping of the generation of note values from a turtle interpretation of Lindenmayer systems. Prusinkiewicz gives an example of a Hilbert curve in which the notes are represented by successive horizontal line segments; the lengths of the segments represents tone durations. The pitches result from the vertical position of the segments and are mapped on the steps of a C major scale (chap. 6 of [13]).

John McCormack [15] compared stochastic processes, Markov models, different variants of generative grammar and Lindenmayer systems in terms of their suitability for musical production. Lindenmayer systems are rewriting systems like generative grammars and generate symbol strings by applying production rules. Originally they were developed to simulate growth processes.

### II(C) CARRY VALUE TRANSFORMATION (CVT): A TOOL FOR FRACTAL FORMATION

In [6] the authors have defined a new transformation named as CVT and shown its use in the formation of fractals.

<u>Definition of CVT in binary number system</u>

If $a=(a_n,a_{n-1},...,a_1)$ and $b=(b_n,b_{n-1},...,b_1)$ are two n-bit strings then $CVT(a,b)=(a_n\wedge b_n,a_{n-1}\wedge b_{n-1},...,a_1\wedge b_1,0)$ is an (n+1) bit string, belonging to the set of non-negative integers, and can be computed bit wise by logical AND operation followed by a 0.

Conceptually, CVT in base-2 number system is same as performing the bit wise XOR operation of the operands (ignoring the carry-in of each stage from the previous stage) and simultaneously the bit wise ANDing of the operands to get a string of carry-bits, the latter string is padded with a ‘0’ on the right to signify that there is no carry-in to the LSB (the overflow bit of this ANDing being always ‘0’ is simply ignored).

<u>Example:</u>

Consider the CVT of the numbers $(13)_{10} \equiv (1101)_2$ and $(14)_{10} \equiv (1110)_2$. Both are 4-bit numbers. The carry value is computed as follows:

<u>Carry: 1 1 0 0 0</u>
Augend: 1 1 0 1
<u>Addend: 1 1 1 0</u>
XOR: 0 0 1 1

Figure 1. Carry genereted in ith column saved in (i-1)th column

In the above example, bit wise XOR gives $(0011)_2 \equiv (3)_{10}$ and bit wise ANDing followed by zero-padding gives $(11000)_2 \equiv (24)_{10}$. Thus $CVT(1101,1110)=11000$ and equivalently in decimal notation one can write $CVT(13,14)=24$. In the next section, CVT is used in different base of the number system in formation of music fractals.

### III(A) GENERATION OF SELF-SIMILAR FRACTALS USING CVT IN DIFFERENT BASES OF THE NUMBER SYSTEMS

A table is constructed that contains only the carry values (or even terms) defined above between all possible integers a’s and b’s arranged in an ascending order of x and y-axis respectively. We observe some interesting patterns in the table. We would like to make it clear how the CV-table is constructed in different bases of number system.

Step 1. Arrange all the integers 0 1 2 3 4 5 6 ... (as long as we want) in ascending order and place it in both, uppermost row and leftmost column in a table.

Step 2. Compute $CVT(a,b)$ as mentioned in II(C) and store it in decimal form in the (a, b) position.

Then we look on the pattern of any integer, and we have made it color. This shows a very beautiful consistent picture, which we see as a fractal as shown below.

Let us do find the fractals in different domain of number system with the help of CVT.

- *Production of Fractal in Binary(2-nary) Number System*

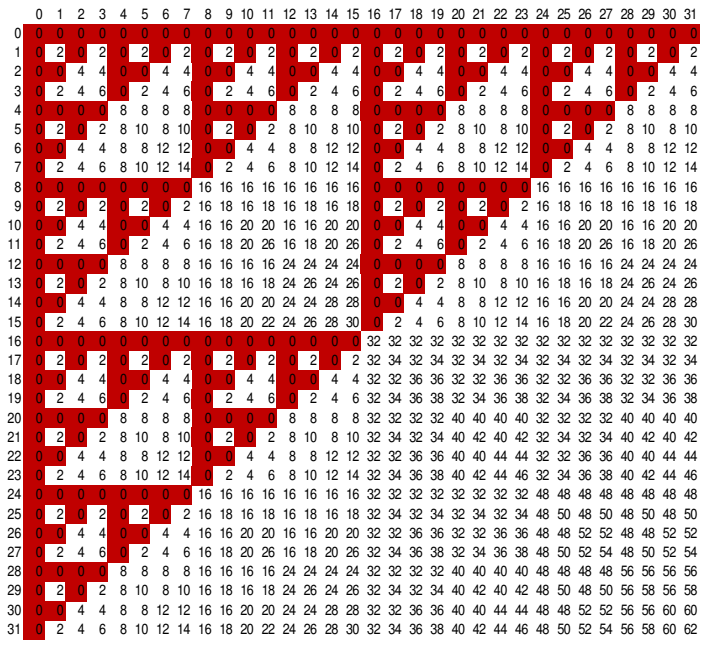

Figure 2. A Fractal Structure on using CVT of Different Integer Values in Binay Number System

*Dimension of this fractal*

For this fractal [see figure 2], number of self-similar copies N=3 and scaling factor S=1/2, where l is the initial length.

So, referring the discussion in II(A), Fractal dimension D is given by

$$3=1/(1/2)^D$$

$$\text{Or } D=\log 3/\log 2 \approx 1.585$$

This is same as the dimension of Sierpinski triangle. Thus CVT fractal as obtained by us can be regarded as a relative to *Sierpinski* triangle [5].

- *CV Table in Ternary Number(3-nary) System*

Here we are applying CVT on the domain of ternary (3-nary) number system and we are having the following table as shown in figure 3. It is mentioned that the procedure can be verbatim copied just by replacing binary by ternary (3-nary). Let us demonstrate how we are computing CVT of two numbers in ternary number system.

Suppose, we want to get CVT of the numbers $(13)_{10} \equiv (111)_3$ and $(14)_{10} \equiv (112)_3$. Both are 3-digit numbers. The carry value is computed as follows:

Carry: 0 0 1 0
Augend: 1 1 1
Addend: 1 1 2
Addition process in ternary: 2 2 0

Therefore, CVT (13,14)= $(0010)_3$ =3.

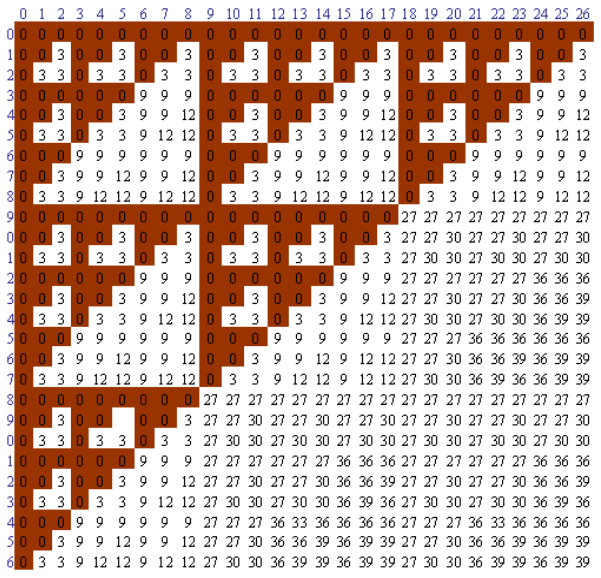

Figure 3. A Fractal Structure on using CVT of Different Integer Values in Ternary Number System

*Dimension of this fractal*

For this fractal [see figure 3], number of self-similar copies N=3 and scaling factor S=1/2, where l is the initial length.

So, referring the discussion in section II, Fractal dimension D is given by

$6=1/(1/3)^D$

Or D= log6/log3 ≈ 1.630929

- *CV Table in 4-nary System*

Here we are applying CVT on the domain of 4-nary number system and we are having the following table. It is mentioned that the procedure can be verbatim copied just by replacing binary by 4-nary, as we have discussed earlier to construct CVT table in binary number system.

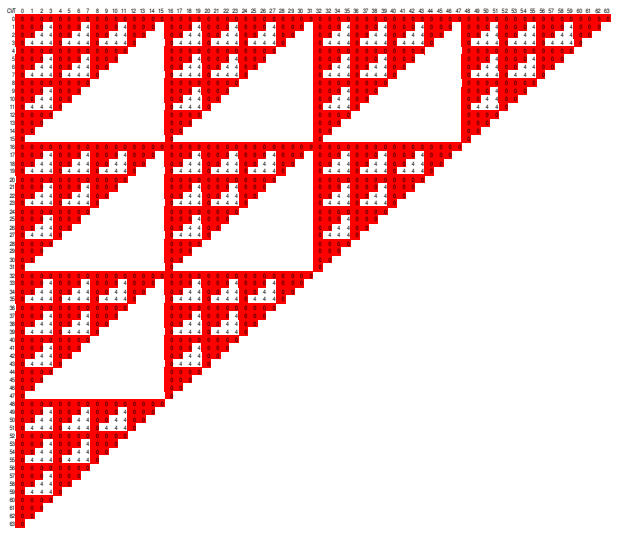

Figure 4. A Fractal Structure on using CVT of Different Integer Values in 4-nary Number System

*Dimension of this pattern*

For this fractal [see figure 4], number of self-similar copies N=3 and scaling factor S=1/2, where l is the initial length.

So, referring the discussion in section II, Fractal dimension D is given by

$10=1/(1/4)^D$

Or D= log10/log4=1.6609

- *CV Table in 5-nary System*

Here we are applying CVT on the domain of 5-nary number system and we are having the following table.

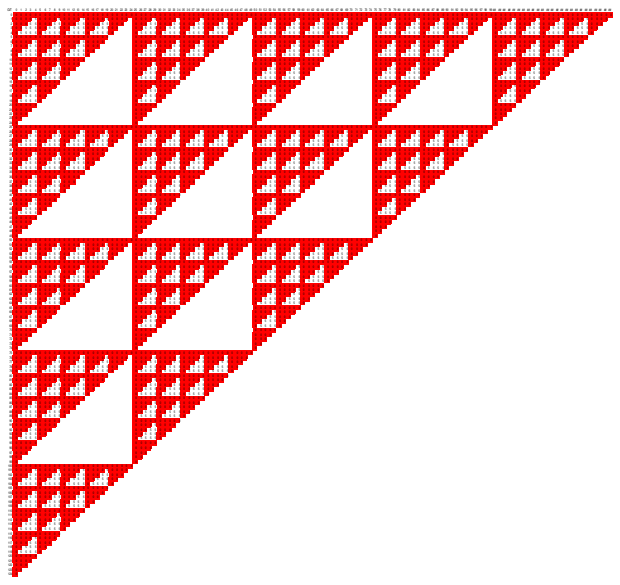

Figure 5. A Fractal Structure on using CVT of Different Integer Values in 5-nary Number System

*Dimension of this pattern*

For this fractal [see figure 5], number of self-similar copies N=3 and scaling factor S=1/2, where l is the initial length.

So, referring the discussion in section II, Fractal dimension D is given by

$15=1/(1/5)^D$

Or D= log15/log5=1.682606

It should be noted that fractal dimension of music (Sri Lankan, Chariots of Fire) is also 1.68 [7]. So this fractal could be a frame of lyrics for the musicians. In general, fractal dimension of arbitrary music is nearly 1.65. And undoubtedly there are a lot of fractals, which are of fractal dimension 1.68, and possibly these fractal-frames make new lyrics for music, which needs further investigation.

Similarly, it could be possible to generate fractals in any base of the number system.

III(B) *Generalization of the concepts in Arbitrary Base of the Number Systems*
Now, we are warmed up to construct the fractal in different number system just following the above procedure. It is also cleared that, when we have calculated the similarity dimension of those fractal in each case we have observed the scaling factor is 1/n and self-similar copies is (1+2+3+…n), for n-nary number system. So let us define our self-similarity dimension formula as follows…

*Here, Number of self − similar copies,*

$$N = \sum_{j=1}^{n} j, \; \& \; Scalling, S = \frac{1}{n}, \; for \; n - nary \; number \; system$$

$$Fractal \; dimension, S_D(n) = \left\{\frac{\log(\sum_{j=1}^{n} j)}{\log n}\right\}$$

$$i.e. S_D(n) = \left\{ log(\frac{n(n+1)}{2}) \Big/ log \; n \right\}$$

*Theorem 1*: The fractal dimension $S_D$ converges to the topological dimension (Euclidian dimension) 2 as the base 'n' of the number system diverges to infinity.

Proof: *Let us try to find the limit when n tends to infinity,*

$$\lim_n S_D(n) = \lim_n \left\{ log(\frac{n(n+1)}{2}) \Big/ log \; n \right\} \left(\frac{\infty}{\infty} \; form\right)$$

$$= \lim \left\{ \left[\frac{2(2n+1)}{2n(n+1)}\right] \Big/ \frac{1}{n} \right\}; \; by \; L'hospital \; rule \; = \lim \left\{\frac{2n^2+n}{n^2+n}\right\} = 2$$

So, starting from the binary number system the fractal dimension of the generated fractal will go on increasing with the increase of the base of the number system and finally it converges to the topological dimension 2.
So far we have discussed we are in position to achieve the fractals having fractal dimension lying in between [1.58, 2), with the help of CVT.

III(C) INCREMENTATION OF FRACTAL DIMENSION OF CVT FRACTALS AND THEIR IMPLICATION IN MUSIC

Let us try to obtain the increment of fractal dimension of the CVT fractals in switching from one base to another base of the number system. In binary number system we have the fractal of fractal dimension 1.58, and in ternary number system we have the fractal of fractal dimension 1.63. So algebraically, it seems the increment of fractal dimension is (1.63-1.58) = 0.05. But it is really not!

To obtain the increment we proceed as follows:

First of all we paste one base (n-1) CVT fractal into base n CVT fractal. Next, the overflowed portion is extracted which can be seen as a self-similar figure. These self-similar pieces derived from overflowed portion would lead to another fractal with some fractal dimension. What we observe, this very measure (fractal dimension) is the actual increment of fractal dimension in switching from one base to another base of the number system.

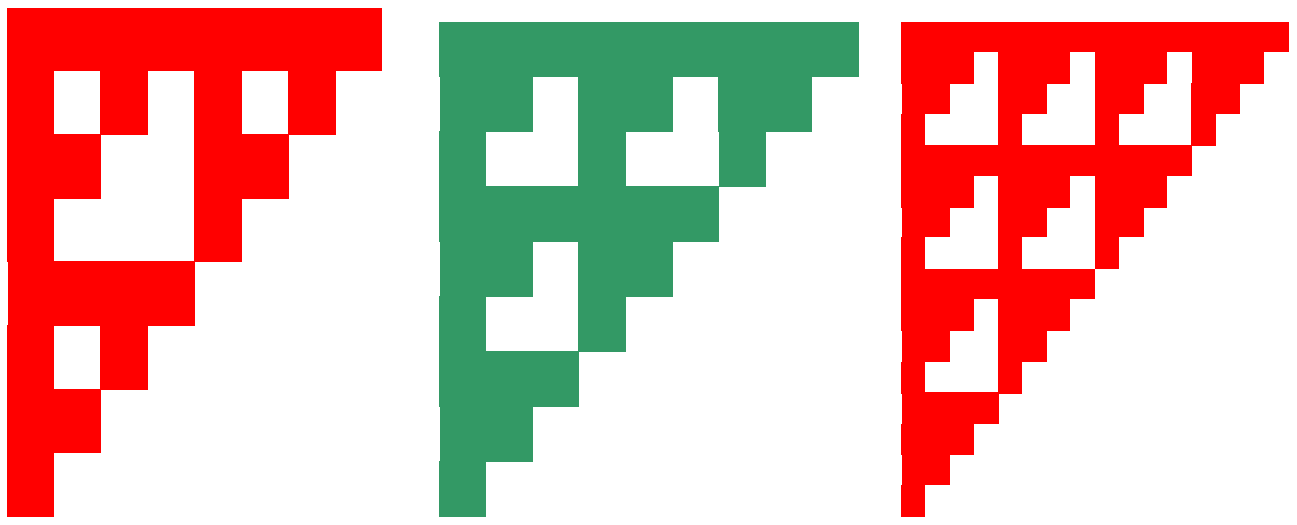

(a) (b) (c)

Figure 6. Shows Binary CVT Fractal generetors in Binary, Ternary, Four-nary form left to right.

1. *Binary Generator over Ternary Generator*

Here we paste binary CVT fractal over ternary CVT fractal [figure 6(a), figure 6(b)]. Next, the overflowed portion can be seen in figure 7, which is a self-similar figure as the CVT fractals were self-similar. These self-similar pieces derived from overflowed portion would lead to another fractal with fractal dimension 1.58 as shown below.

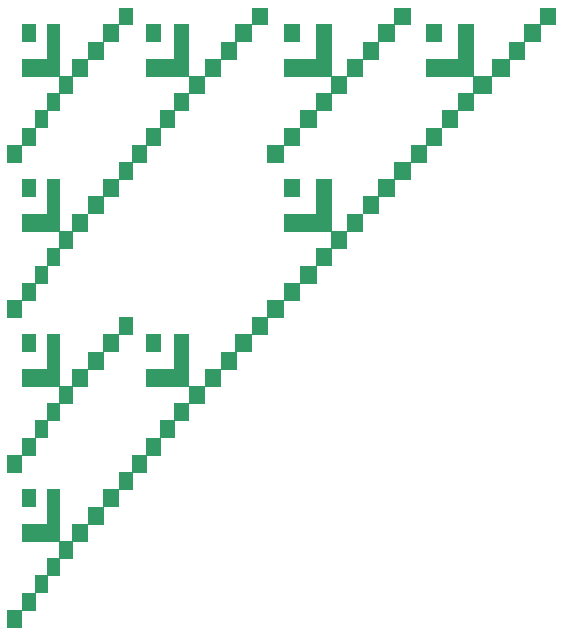

Figure 7. Binary CVT generetor over Ternary generetor lead to the another fractal.

This generator leads to another fractal; let us calculate the fractal dimension of the generated fractal as follows:

*Dimension of this fractal*
For this fractal, N=3, S=1/2, where l is the initial length.

Fractal dimension D is given by: $3=1/(1\ /\ 2)^D$

Or $D=\log 3/\log 2 \approx 1.585$

This is same as the dimension of Sierpinski triangle. Thus CVT fractal as obtained by us can be regarded as a relative to *Sierpinski* triangle.

Therefore, we could say that the amount of increment is likely 1.58.

2. *Ternary Generator over Four-nary Generator*

Here we paste ternary CVT fractal over 4-nary CVT fractal [figure 6(b), figure 6(c)]. Next, the overflowed portion can be seen in figure 8, which is a self-similar figure as the CVT fractals were self-similar. These self-similar pieces derived from overflowed portion would lead to another fractal with fractal dimension 1.58 as shown below.

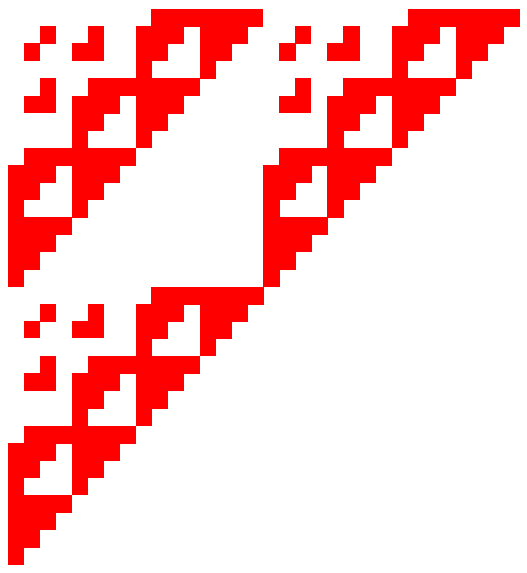

Figure 8. Ternary CVT generetor over Four-nary generetor leading to another fractal.

This generator also leads to another fractal whose similarity dimension is also ≈ 1.585. Thus, CVT fractal as obtained in this process can be regarded as a relative to *Sierpinski* triangle. Here also we are having the same increment as above. In fact, in general we could make a conjecture that if we like to paste an k-nary CVT generator over an (k+1)-nary CVT generator, then we will be able to have another generator which lead to the fractal of fractal dimension 1.58, i.e. the attractor fractal is always homeomorphic to Sierpinski Gasket.

Let us consider two music of fractal dimension 1.66 and 1.68 respectively. And music is nothing but a collection of notes in different orientation. And we know that, the fractal dimension is a measure of amount of chaos or amount of information. So in this regard we can say the music of fractal dimension 1.68 contains more information about musical notes than the music of fractal dimension 1.66. It is noticeable that, these two music are not too much different according to fractal dimension. Now question is how much more information is there in the music of fractal dimension of 1.68 than 1.66. We are claiming this amount of information is 1.58. It is the firm conviction of the authors that the music fractals thus generated will have this kind of incremental fractal dimensions with regards to certain base or bases. So immediately we will take up this investigation as our future research effort.

## IV. Conclusion and Future Research directions

A function is being used to construct different fractals and these fractals could be taken as a new texture of lyrics to the lyricists. We conjecture that the switching from one music fractal to another is nothing but enhancing a constant amount of fractal dimension. This constant measure of fractal dimension has to be correlated with the various possibilities of musical notes in different orientations. In the coming future, a major thrust would be given in this direction.